\documentclass[aps,prl,reprint,superscriptaddress,floatfix,amsmath]{revtex4-1}
\usepackage{graphicx,epstopdf}
\begin{document}

\title{First direct determination of the superallowed $\beta$-decay $Q_{EC}$-value for $^{14}$O }
%\title{First direct determination of the $Q_{EC}$-value for $^{14}$O, a superallowed $\beta$ emitter, to test the Standard Model.}
\author{A.A. Valverde}
\email[]{valverde@nscl.msu.edu}
\affiliation{National Superconducting Cyclotron Laboratory, East Lansing, Michigan 48824 USA}
\affiliation{Department of Physics and Astronomy, Michigan State University, East Lansing, Michigan 48824, USA}

\author{G. Bollen}
\affiliation{Department of Physics and Astronomy, Michigan State University, East Lansing, Michigan 48824, USA}
\affiliation{Facility for Rare Isotope Beams, East Lansing, Michigan 48824 USA}

\author{M. Brodeur}
\affiliation{Department of Physics, University of Notre Dame, Notre Dame, Indiana 46556, USA}

\author{R.A. Bryce}
\affiliation{Central Michigan University, Mount Pleasant, Michigan 48859, USA}

\author{K. Cooper}
\affiliation{National Superconducting Cyclotron Laboratory, East Lansing, Michigan 48824 USA}
\affiliation{Department of Chemistry, Michigan State University, East Lansing, Michigan 48824, USA}

\author{M. Eibach}
\affiliation{National Superconducting Cyclotron Laboratory, East Lansing, Michigan 48824 USA}

\author{K. Gulyuz}
\affiliation{National Superconducting Cyclotron Laboratory, East Lansing, Michigan 48824 USA}

\author{C. Izzo}
\affiliation{National Superconducting Cyclotron Laboratory, East Lansing, Michigan 48824 USA}
\affiliation{Department of Physics and Astronomy, Michigan State University, East Lansing, Michigan 48824, USA}

\author{D.J. Morrissey}
\affiliation{National Superconducting Cyclotron Laboratory, East Lansing, Michigan 48824 USA}
\affiliation{Department of Chemistry, Michigan State University, East Lansing, Michigan 48824, USA}

\author{M. Redshaw}
\affiliation{National Superconducting Cyclotron Laboratory, East Lansing, Michigan 48824 USA}
\affiliation{Central Michigan University, Mount Pleasant, Michigan 48859, USA}

\author{R. Ringle}
\affiliation{National Superconducting Cyclotron Laboratory, East Lansing, Michigan 48824 USA}

\author{R. Sandler}
\affiliation{National Superconducting Cyclotron Laboratory, East Lansing, Michigan 48824 USA}
\affiliation{Department of Physics and Astronomy, Michigan State University, East Lansing, Michigan 48824, USA}

\author{S. Schwarz}
\affiliation{National Superconducting Cyclotron Laboratory, East Lansing, Michigan 48824 USA}

\author{C.S. Sumithrarachchi}
\affiliation{National Superconducting Cyclotron Laboratory, East Lansing, Michigan 48824 USA}

\author{A.C.C. Villari}
\affiliation{Facility for Rare Isotope Beams, East Lansing, Michigan 48824 USA}

\date{\today}

\begin{abstract}
We report the first direct measurement of the $^{14}\text{O}$ superallowed Fermi $\beta$-decay $Q_{EC}$-value, the last of the so-called ``traditional nine'' superallowed Fermi $\beta$-decays to be measured with Penning trap mass spectrometry. $^{14}$O, along with the other low-$Z$ superallowed $\beta$-emitter, $^{10}$C, is crucial for setting limits on the existence of possible scalar currents. The new ground state $Q_{EC}$ value, 5144.364(25) keV, when combined with the energy of the $0^+$ daughter state, $E_x(0^+)=2312.798(11)$~keV [Nucl. Phys. A {\bf{523}}, 1 (1991)], provides a new determination of the superallowed $\beta$-decay $Q_{EC}$ value, $Q_{EC}(\text{sa}) = 2831.566(28)$ keV, with an order of magnitude improvement in precision, and a similar improvement to the calculated statistical rate function $f$. This is used to calculate an improved $\mathcal{F}t$-value of 3073.8(2.8) s.
\end{abstract}

\maketitle

The study of superallowed Fermi $\beta$-decays, those occurring between nuclear analogue states with nuclear spin and parity $J^{\pi}=0^+$ and isospin $T=1$, provide a powerful tool for testing the foundation of the standard model of the electroweak interaction. An important feature of these decays is that they can provide a test of the conserved-vector-current (CVC) hypothesis and set limits on the existence of scalar currents. Furthermore, the decay energies are employed for the most stringent test of the unitarity of the Cabibbo-Kobayashi-Maskawa (CKM) matrix, by their contribution to the value of $V_{ud}$, used in investigations of physics beyond the standard model \cite{Hardy14}.

The CVC hypothesis asserts that the weak vector coupling constant, $G_V$, is not renormalized in the nuclear medium. As such, the experimental $ft$-value for superallowed Fermi $\beta$-decay should be the same for all such transitions, independent of the nucleus. However, small modifications must be included to account for radiative corrections and the fact that isospin is not an exact symmetry. Hence, the corrected $\mathcal{F}t$-value should be the same for all superallowed $\beta$-decays:
\begin{equation}
\mathcal{F}t=ft(1+\delta'_r)(1+\delta_{\text{NS}}-\delta_c)=\frac{K}{2G_V ^2(1+\Delta_R^V)}, \label{eq:FT}
\end{equation}
where $\Delta_R^V$ is the transition-independent part of the radiative correction, $\delta_r'$ and $\delta_{\text{NS}}$ are the transition-dependent parts of the radiative correction, and $\delta_c$ is the isospin-symmetry-breaking correction. 
The corrected $\mathcal{F}t$-value can also be expressed in terms of a constant $K$ ($K/(\hbar c)^6=8120.2787(11)\times10^{-10}\text{ GeV}^{-4}\times\text{s}$), $\Delta_R^V$, and $G_V$, the weak vector coupling constant. The value for $G_V$ determined from the $\mathcal{F}t$-value can, when combined with $G_F$, the weak-interaction constant determined from purely leptonic muon decays \cite{Hardy14}, be used to calculate $V_{ud}$ and thus provide a test of the unitarity of the CKM matrix. 

The $ft$-value is the product of the statistical rate function, $f$, and the partial half-life of the decay, $t$. Three experimental quantities contribute to its determination:  the half-life of the parent state $t_{1/2}$ and the branching ratio of the $0^+ \to 0^+$ $\beta$-decay, which are used to obtain $t$, and the decay transition energy $Q_{EC}$, which is used to calculate $f$ \cite{Hardy14}. $f$ depends on the fifth power of the $Q_{EC}$-value \cite{Hardy15}, hence, it is essential to have a precise and accurate determination of $Q_{EC}$.  

On-line Penning trap mass spectrometry has provided  significant contributions to the compilation of $\mathcal{F}t$ data by providing high-precision direct $Q_{EC}$-value measurements \cite{Hardy05}. Of the ``traditional nine'' superallowed $\beta$-decay isotopes, those with stable daughter nuclei, $^{10}$C \cite{Eronen11,Kwiatkowski13}, $^{26}$Al$^m$ \cite{Eronen06,George08}, $^{34}$Cl \cite{Eronen09}, $^{38}$K$^m$ \cite{Eronen09}, $^{42}$Sc \cite{Eronen06}, $^{46}$V \cite{Savard05,Eronen06,Eronen11}, $^{50}$Mn \cite{Eronen08} and $^{54}$Co \cite{Eronen08} have all had their $Q$-values determined with a Penning trap; only $^{14}$O remains to be determined, despite multiple attempts at other facilities.

Low-$Z$, superallowed $\beta$-emitters like $^{14}$O are particularly significant for setting limits on the existence of scalar currents.  While the CVC hypothesis states that $\mathcal{F}t$ should be the same for all superallowed $0^+ \to 0^+$ $\beta$-decays, if there is a scalar interaction, an additional term approximately inversely proportional to $Q_{EC}$ would be present  in $\mathcal{F}t$. As $Q_{EC}$-values are smaller for lower-$Z$ isotopes, these isotopes would show the largest deviation in $\mathcal{F}t$ from a constant value.

The $Q_{EC}$-value of $^{14}\text{O}$ given in the current compilation of $\mathcal{F}t$ data \cite{Hardy14} is determined from a weighted average of threshold energy measurements of the $^{12}\text{C}(^{3}\text{He},n)^{14}\text{O}$, $^{12}\text{C}(^{3}\text{He},p)^{14}\text{N}$, and $^{14}\text{N}(p,n)^{14}\text{O}$\cite{Butler61,Bardin62,Roush70,White77, Tolich03} reactions. This ensemble of data has one of the largest statistical spreads among all $T=1$ transitions, resulting in a reduced-$\chi^2$ of 4.4. Therefore, the uncertainty has been inflated by a factor of 2.1 to 280 eV in its evaluation \cite{Hardy14}. Previous comparisons of reaction threshold energy measurements to Penning trap measurements for other isotopes have shown as much as a 6$\sigma$ difference \cite{Savard05,Eronen06,Eronen08}. Hence, a direct measurement of the $Q_{EC}$-value of the $^{14}$O superallowed Fermi $\beta$-decay is necessary.

\begin{figure}[t!]
 \includegraphics[width=\columnwidth]{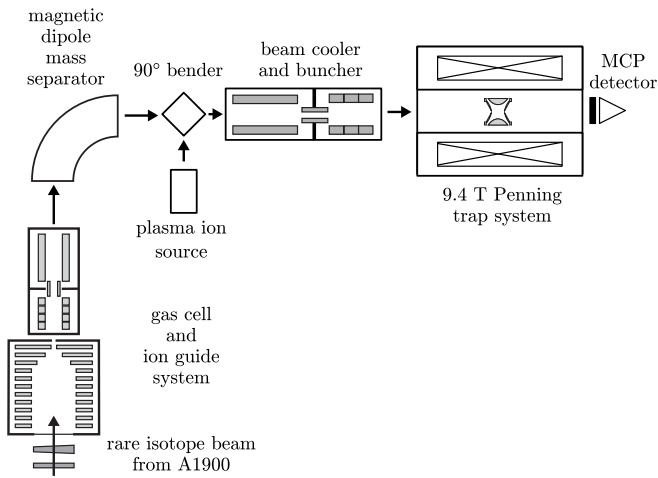}
 \caption{A schematic diagram showing the major elements of the gas cell and LEBIT facility.\label{fig:schema}}
\end{figure}

In this Letter, we report on the first direct ground state $\beta$-decay $Q_{EC}$-value measurement of $^{14}$O, carried out at the National Superconducting Cyclotron Laboratory (NSCL) using the 9.4T Penning trap mass spectrometer at the Low-Energy Beam and Ion Trap (LEBIT) facility\cite{Ringle13}. The LEBIT facility is unique among Penning trap mass spectrometry facilities in being able to perform high-precision measurements on rare isotopes produced by projectile fragmentation. A schematic diagram of the gas cell and LEBIT facility is shown in Fig. \ref{fig:schema}. The radioactive $^{14}$O was produced by impinging a 150 MeV/$u$ primary beam of $^{16}$O on a 2233 $\mu$m thick beryllium target at the Coupled Cyclotron Facility at the NSCL. The products were passed through the A1900 fragment separator \cite{Morrissey03}, to separate the secondary beam, consisting of fully-stripped $^{14}\text{O}$  $(46\%)$ , $^{13}\text{N}$  $(36\%)$, and $^{12}\text{C}$  $(13\%)$ ions.

The secondary beam then passed through the momentum compression beamline, where it was decelerated with aluminum degraders at several points before passing through a fused silica wedge. The thickness of the degraders was selected to stop the $^{13}\text{N}$; the remainder of the beam then entered the gas cell with an energy of less than 1 MeV/$u$ \cite{Cooper14}. The ions in the gas cell were stopped by collisions with high-purity helium gas at a pressure of about 93 mbar; during this process, the highly-charged ions recombined to their singly-charged state. These ions were transported  in the gas cell by a combination of RF and DC fields as well as gas flow. They were then extracted into a radiofrequency quadrupole (RFQ) ion-guide and transported through a magnetic dipole mass separator with a resolving power greater than $500$, which was used to separate $^{14}\text{O}^+$ from the other contaminants in the beam. A beta detector located after the mass separator was used to confirm the identity of the $^{14}\text{O}$ by measuring the average half-life of the isotope. The measured $t_{1/2}=70.2(2.6)$ s is consistent with a recent more-precise measurement for $^{14}\text{O}$, $t_{1/2}=70.619(11)$~s \cite{Laffoley13}. Following the mass separator, the ions were transported to the LEBIT facility.

In the LEBIT facility, the $^{14}\text{O}^+$ ions first entered a cooler-buncher, a two-staged helium-gas-filled RFQ ion trap \cite{Schwarz03}. In the first stage, moderate pressure helium gas was used to cool the ions in a large diameter RFQ ion guide; in the second, the ions were accumulated, cooled, and released to the LEBIT Penning trap in pulses of approximately 100 ns \cite{Ringle09}. To further purify the beam, a fast kicker in the beam line between the cooler-buncher and the Penning trap was used as a time-of-flight mass gate.

The LEBIT 9.4T Penning trap features a high-precision hyperbolic electrode system inside a 9.4T actively-shielded magnet \cite{Ringle13}. Retardation electrodes upstream of the Penning trap were used to decelerate the ion pulses to low energy before entering the trap. The last section of these electrodes are quadrisected radially to form a ``Lorentz steerer'' \cite{Ringle07b} that forces the ion to enter the trap off-axis and to perform a magnetron motion with frequency $\nu_-$ once the trapping potential is applied.

After capture, the trapped ions were purified using dipole cleaning. The ions to be cleaned, $^{14}\text{N}^+$ and $^{12}\text{C}^1\text{H}_2^+$, were identified and then excited using azimuthal RF dipole fields at their reduced cyclotron frequency $\nu_+$ and driven to a large enough radius that they did not interfere with the measurement. The time-of-flight cyclotron resonance technique (TOF-ICR) \cite{Bollen90,Konig95} was then used to determine the ions' cyclotron frequency. In the TOF-ICR technique, the ions are excited with an azimuthal quadrupole RF field with $\nu_{RF}\approx\nu_c=\nu_++\nu_-$. When $\nu_{RF}=\nu_c$, there is a periodic conversion between the initial magnetron motion and reduced cyclotron motion. The conversion process results in a significant change in the ions' radial energy, which can be detected when ions are ejected from the trap and the radial energy is converted to axial energy that is measured via the time of flight to a microchannel plate (MCP) detector. Maximal change in radial energy, and thus in time of flight, occurs when $\nu_{RF}=\nu_c$, so the time-of-flight resonance can be built by measuring multiple bunches of ions and varying $\nu_{RF}$ around $\nu_c$. From a fit of the theoretical line shape to this resonance one can determine the cyclotron frequency $\omega_c=2 \pi \nu_c = q/m \cdot B$, where $m$ is the mass, $q$ is the charge,  and $B$ is the magnetic field strength.

\begin{figure}[t!]
\includegraphics[width=\columnwidth]{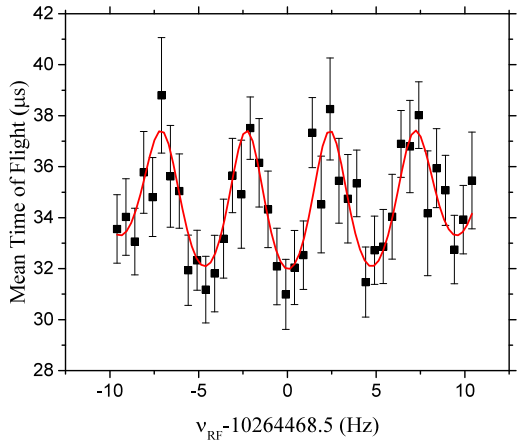}
 \caption{(color online). A sample $^{14}\text{O}^+$ time-of-flight  cyclotron resonance used for the determination of the frequency ratio of $\nu_{\text{ref}}^{\text{int}}(^{14}\text{N}^+)/\nu_c(^{14}\text{O}^+)$. The solid red curve represents a fit of the theoretical profile \cite{Kretzschmar07}.\label{fig:O14}}
\end{figure}

\begin{figure}[b!]
\includegraphics[width=\columnwidth]{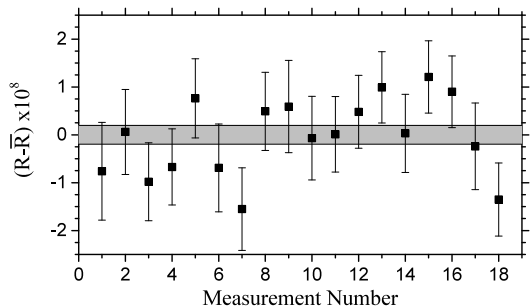}
 \caption{Measured cyclotron frequency ratios $R=\nu_{\text{ref}}^{\text{int}}(^{14}\text{N}^+)/\nu_c(^{14}\text{O}^+)$ relative to the average value $\bar{R}$; the grey bar represents the $1\sigma$ uncertainty in $\bar{R}$.\label{fig:O14s}}
\end{figure}

In these measurements, a 250-ms Ramsey excitation scheme \cite{Bollen92b,George07} was used; a 50-ms burst of RF was applied, followed by a 150-ms waiting period, and then a final 50-ms burst of RF.  The Ramsey resonances were fitted to the theoretical line shape \cite{Kretzschmar07}, and the cyclotron frequency was thus determined; a sample resonance for $^{14}\text{O}^+$ can be seen in Fig. \ref{fig:O14}. Measurements of the reference ion $^{14}\text{N}^+$ cyclotron frequency were conducted between measurements of the $^{14}\text{O}^+$ cyclotron frequency. $^{14}\text{N}^+$ ions were produced in a plasma ion source located upstream of the cooler-buncher (see Fig. \ref{fig:schema}) and otherwise treated the same as the $^{14}\text{O}^+$. The experimental result is the cyclotron frequency ratio $R=\nu_{\text{ref}}^{\text{int}}(^{14}\text{N}^+)/\nu_c(^{14}\text{O}^+)$, where $\nu_c^{\text{int}}(^{14}\text{N}^+)$ is the interpolated cyclotron frequency from the $^{14}\text{N}^+$ measurements bracketing each  $^{14}\text{O}^+$ measurement. A series of eighteen measurements of the $^{14}\text{O}^+$ cyclotron frequency were taken over a 24-hour period. The weighted average of these measurements is $\bar{R}=$1.000 394 407 5 (19), where the statistical uncertainty is given; the individual values of $R$ relative to the average $\bar{R}$ can be seen in Fig. \ref{fig:O14s}. 

Previous work has shown that the effect of nonlinear magnetic field fluctuations on the ratio $R$ should be less than $1\times 10^{-9}$ per hour, which was our measurement time \cite{Ringle07}. The presence of isobaric contaminants in the trap during a measurement could lead to a systematic frequency shift \cite{Bollen92}; this effect was minimized by ensuring no contaminants were present at a level exceeding a few percent and by limiting the total number of ions in the trap. This was done by only analyzing events with five or fewer ions detected; as our measured MCP efficiency is 63\%, this corresponds to eight or fewer ions in the trap. As $^{14}\text{O}^+$ and $^{14}\text{N}^+$ form an isobaric doublet, most of the systematic effects, such as relativistic shifts in the measured masses due to differences in velocity and additional shifts due to trap field imperfections and special relativity caused by differences in orbital radius in the trap are negligible, as the masses and orbital radii are practically identical. The specific shifts to the ratio due to a possible mismatch in the radii are calculated \cite{Brown86} to be less than $4\times10^{-10}$,which is negligible compared to the statistical uncertainty. The Birge ratio \cite{Birge32} for the measurement was 1.02(11), which indicates that the fluctuations are statistical in nature. 
\begin{figure}[b!]
\includegraphics[width=\columnwidth]{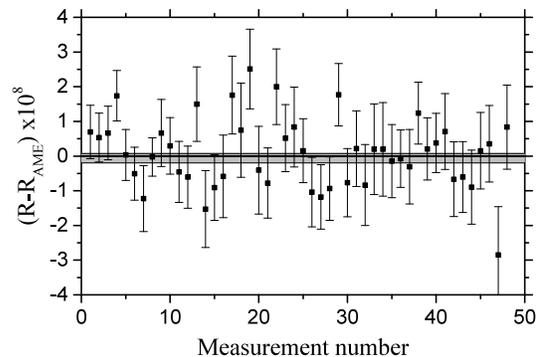}
 \caption{Measured ratios $R=\nu_{\text{ref}}^{\text{int}}(^{14}\text{N}^+)/\nu_c(^{12}\text{C}^1\text{H}_2^+)$ of $\text{CH}_2$ relative to the value calculated from the 2012 atomic mass evaluation \cite{AME12}. The grey bar represents the weighted average measured value $\bar{R}$ and its uncertainty; the uncertainty of the \textsc{Ame2012} value, $1.7\times10^{-11}$, is not visible on this graph.\label{fig:CH2s}}
\end{figure}

The possibility of additional unknown systematic effects was investigated through a measurement of the ratio $R$ of a second mass doublet, $R=\nu_{\text{ref}}^{\text{int}}(^{14}\text{N}^+)/\nu_c(^{12}\text{C}^1\text{H}_2^+)$. This value can also be calculated as $R= m(^{12}\text{C}^1\text{H}_2^+)/m(^{14}\text{N}^+)$ using the \textsc{Ame2012} masses of $^{14}$N, $^{12}$C, $^1$H and the electron, the first ionization energies of $\text{CH}_2$ and nitrogen, and the heat of formation of  $\text{CH}_2$ \cite{AME12,Willitsch02,NIST-AST,NIST-WB}. The measured value agrees with the literature value to within $6.0\times10^{-10}$, with an uncertainty of $1.4\times10^{-9}$, as shown in Fig. \ref{fig:CH2s}. Thus, any mass dependent shifts in the present measurement of a doublet at mass 14 would be smaller than the statistical uncertainty in the $^{14}\text{O}$ measurement.

The $^{14}\text{O}$ ground state $Q_{EC}$-value and atomic mass can then be calculated from:
\begin{align}
Q_{EC}(\text{g.s.})&= \left[(M(^{14}\text{N})-m_e)c^2+b_{\text{N}}\right][\bar{R}-1]+b_N-b_O\label{eq:Qec}\\
\intertext{and:}
M(^{14}\text{O})&= \left[M(^{14}\text{N})-m_e+\frac{b_{\text{N}}}{c^2}\right]\bar{R}+m_e-\frac{b_{\text{O}}}{c^2}, \label{eq:mass}
\end{align}
respectively, where $m_e$ is the electron mass, $b_{\text{N}}$ and $b_{\text{O}}$ are the first ionization energy of nitrogen and oxygen, respectively,  and the \textsc{Ame2012} mass of $^{14}\text{N}$ is used \cite{NIST-AST,AME12}. The resulting ground state $Q_{EC}$-value and mass excess are reported in Table \ref{tab:meas}, as well as the currently-accepted values. The new ground state $Q_{EC}$-value measurement agrees well with the accepted values, but is an order of magnitude more precise.  The new mass excess differs by 3--$\sigma$ from the \textsc{Ame2012} value, which is based on threshold energy measurements of the  $^{14}\text{N}(p,n)^{14}\text{O}$ and $^{26}\text{Mg}(^{3}\text{He},t)^{26}\text{Al}-^{14}\text{N}(^{3}\text{He},t)^{14}\text{O}$ reactions \cite{AME12}.
\begin{table}[t!]
 \caption{\label{tab:meas}A comparison of the measured ground state $Q_{EC}$-value and mass excess with the 2014 critical survey of superallowed $0^+\to0^+$ nuclear $\beta$-decays \cite{Hardy14} and the \textsc{Ame2012} atomic mass evaluation \cite{AME12}.}
 \begin{ruledtabular}
 \begin{tabular}{c c c}
Measurement & Value (keV)  & Ref.\\
\hline
$Q_{EC}$(g.s.) &5144.32(28) & \cite{Hardy14}\\
$Q_{EC}$(g.s.) &5144.364(25)& This work\\
ME($^{14}$O)& 8007.457(110) & \cite{AME12}\\
ME($^{14}$O)& 8007.781(25) & This work\\
\hline
\end{tabular}
\end{ruledtabular}
\end{table}

The $\mathcal{F}t$-value was calculated using Eq. \eqref{eq:FT}. The values for partial half-life $t$ (calculated from the average measured half-life $t_{1/2}$ and the branching ratio) and theoretical corrections $\delta'_r$ and $\delta_c-\delta_{\text{NS}}$ were taken from \cite{Laffoley13,Hardy14}. The superallowed $Q_{EC}(\text{sa})$-value can then be calculated from $Q_{EC}(\text{sa})=Q_{EC}(\text{g.s.})-E_x(0^+)$, where $E_x(0^+)=2312.798(11)$ keV\cite{AjzenbergSelove91} is the energy of the $0^+$ daughter state in $^{14}\text{N}$. The statistical rate function $f$ was calculated with $Q_{EC}(\text{sa})$ and the parametrization presented in \cite{Hardy15}. Compared with the value published in \cite{Hardy14}, the uncertainty in $f$ is reduced by an order of magnitude. The uncertainty in our calculated $\mathcal{F}t$-value is also reduced compared to \cite{Hardy14}. The contribution of the $Q_{EC}$-value to the uncertainty in $\mathcal{F}t$ is now 0.2 s, compared to the 1.7 s in \cite{Hardy14}. These results are presented in Table \ref{tab:calcs}.
\begin{table}[t!]
 \caption{\label{tab:calcs}A comparison of the calculated superallowed $Q_{EC}$-value, statistical rate function $f$ and the corrected $\mathcal{F}t$-value with the 2014 critical survey of superallowed $0^+\to0^+$ nuclear $\beta$-decays \cite{Hardy14}.}
 \begin{ruledtabular}
 \begin{tabular}{c c c}
Calculation & Value & Ref.\\
\hline
$Q_{EC}(\text{sa})$ & 2831.23(23) keV & \cite{Hardy14}\\
$Q_{EC}(\text{sa})$ & 2831.566(28) keV & This work\\
$f$& 42.771(23) & \cite{Hardy14}\\
$f$& 42.8055(28) & This work\\
$\mathcal{F}t$ &3071.4(3.2) s& \cite{Hardy14}\\
$\mathcal{F}t$ &3073.8(2.8) s& This work\\
\hline
\end{tabular}
\end{ruledtabular}
\end{table}

\begin{figure}[t!]
\includegraphics[width=\columnwidth]{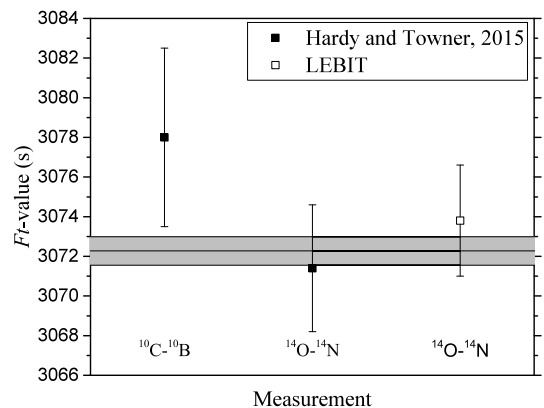}
 \caption{Comparisons of the $^{10}$C and $^{14}$O $\mathcal{F}t$-values from \cite{Hardy14}, shown with a solid square, and the new value calculated in this work, shown as a hollow square, with the average $\mathcal{F}t$-value from \cite{Hardy14} and its uncertainty, represented as a grey line.\label{fig:Fts}}
\end{figure}

In summary, the first direct determination of the $Q_{EC}$-value of the last remaining of the ``traditional nine'' superallowed Fermi transitions was carried out to a precision of 25 eV, making it the most precisely-known $Q_{EC}$-value for determining an $\mathcal{F}t$-value used in testing the CVC hypothesis. The uncertainty in $\mathcal{F}t$ is now dominated by the uncertainty in the experimentally-measured branching ratio, and the theoretical correction $\delta_c-\delta_{\text{NS}}$. The increase in $Q_{EC}$ value results in a slight increase to the $\mathcal{F}t$-value; while this does not raise it outside the error bars of the average $\mathcal{F}t$-value for superallowed transitions, the direction of the shift is consistent with the shift in $\mathcal{F}t$-value observed for $^{10}$C after its $Q_{EC}(\text{g.s.})$-value was determined via Penning trap mass spectrometry \cite{Eronen11,Kwiatkowski13}, which could indicate the presence of a scalar current \cite{Hardy14}, as shown in Fig. \ref{fig:Fts}. In order to confirm or refute this, an improved measurement of the branching ratios for both nuclei is needed.

This work was conducted with the support of Michigan State University and the National Science Foundation under Contracts No. PHY-1102511 and PHY-1307233.

\end{document}